\documentstyle{article}
\begin{document}
\newtheorem{theorem}{Theorem}
\newtheorem{lemma}{Lemma}

\title{A causal statistical family of dissipative divergence type fluids}

\author{Oscar A.  Reula\thanks{Researcher of CONICET}, and Gabriel B.
Nagy\thanks{Fellow of CONICOR}}

\maketitle


\begin{center}
Facultad de Matem\'atica Astronom\'{\i}a y F\'{\i}sica \\
Universidad Nacional de C\'ordoba \\
Dr. Medina Allende y Haya de la Torre, \\
Ciudad Universitaria, \\
(5000) C\'ordoba, Argentina.
\end{center}

\begin{abstract}
In this paper we investigate some  properties, including causality, of
a particular class of relativistic dissipative fluid theories of
divergence type. This set is defined as those theories coming from a
statistical description of matter, in the sense that the three tensor
fields appearing in the theory can be expressed as the three first
momenta of a suitable distribution function. In this set of theories
the causality condition for the resulting system of hyperbolic partial
differential equations is very simple and allow to identify a subclass
of manifestly causal theories, which are so for all states outside
equilibrium for which the theory preserves this statistical
interpretation condition. This subclass includes the usual equilibrium
distributions, namely Boltzmann, Bose or Fermi distributions, according
to the statistics used, suitably generalized outside equilibrium.
Therefore this gives a simple proof that they are causal in a
neighborhood of equilibrium.  We also find a bigger set of dissipative
divergence type theories which are only pseudo-statistical, in the
sense that the third rank tensor of the fluid theory has the symmetry
and trace properties of a third momentum of an statistical
distribution, but the energy-momentum tensor, while having the form of
a second momentum distribution, it is so for a different distribution
function.  This set also contains a subclass (including the one already
mentioned) of manifestly causal theories.
\end{abstract}

\section{Introduction}

In the last few years there has been a considerable effort to
understand dissipation in relativistic theories of fluids.  Straight
forward generalizations to relativity of the Navier-Stokes scheme
\cite{e} resulted in systems with an ill-posed initial value
formulation\cite{hl}.  Even if these generalizations would have worked,
they would have yielded a parabolic system, well-posed in the
mathematical sense, but unacceptable on physical grounds due to the
presense of infinite propagation velocities.  Thus, alternative
theories were proposed resulting in a formalism having an extended
number of dynamical variables and where dissipation is at the
microscopic level completely different to the standard parabolic
dissipation of Navier-Stokes equations, but which behaves at measurable
scales in the same way \cite{g1}, \cite{g2}. Due to the problems
encountered on the generalizations alluded above, one of the basic
requirements to be checked on these alternative theories was their
relativistic causality \cite{gl1}, \cite{gl2}.

Among the alternative theories, we specialize to the case of theories
of divergence type \cite{gl1}, \cite{p}, that is theories with the
following structure
\begin{eqnarray}
\nabla_a N^a &=& 0 \label{def1} \\
\nabla_a T^{ab} &=& 0  \label{def2} \\
\nabla_a A^{abc} &=& I^{bc} \label{def3}
\end{eqnarray}
where the tensors $A^{abc}$ (tensor of fluxes) and $I^{bc}$
(dissipation-source tensor) are supposed to be functions on a smaller
set of fields, which in principle can be taken to be the direct
observables of the theory, namely the particle number $N^a$ and the
energy-momentum $T^{ab}$ of the fluid. The tensor $I^{ab}$ represents
the non-equilibrium interactions in the fluid, and vanishes at the
local equilibrium fluid configurations, i.e. at the momentarily static
configurations.  We focus on this particular type of theories, not just
because their structure is simpler, but also because:  a) They already
contain enought degrees of freedom to account for the usual description
of nonrelativistic dissipative fluids. b) Since these theories
generically develop shock waves, it is desiderable their structure has
this form in order for the shock wave solutions to make sense.

The requirement of an {\it entropy law}, that is the existence of an
extra vector field whose divergence, by the sole virtue of the above
equations, is a function of the basic fields at the point (and not of
any of its derivatives), puts severe restrictions in the theory.  In
particular it implies the existence of a single generating function
$\chi$, and variables $(\zeta, \zeta^a, \zeta^{ab})$, with $\zeta$
negative, $\zeta^a$ future directed timelike, and $\zeta^{ab}$
symmetric and trace free,
such that
\begin{eqnarray}
N_a &=& \frac{\partial^2 \chi}{\partial \zeta \partial \zeta^a}
\label{defn} \\
T_{ab} &=& \frac{\partial^2 \chi}{\partial \zeta^a \partial \zeta^b}
\label{deft} \\
A_{abc} &=& \frac{\partial^2 \chi}{\partial \zeta^a \partial \zeta^{bc}}
\label{defa}
\end{eqnarray}
That is, this generating function determines completely the principal part,
in the sense of the theory of partial differential equations,
of the above system, and so its causal properties.

Non-dissipative fluids can also be described with divergence type
theories by simply choosing a generating function of the form
$\chi(\zeta,\zeta^a)$, and setting to zero the tensor $I^{ab}$.  This
kind of generating function describes a perfect fluid theory, with
$\chi$ related with the fluid's equation of state, as explained in
section II and Appendix A.

In this paper we first define a subclass of non-dissipative divergence
type theories  that we call {\it statistical}. This perfect fluid
theories can be characterized as those having a statistical origin, in
the sense that the particle-number current and the stress-energy tensor
can be expressed as suitable linear combination of the first two
momenta of some distribution function. This requirement singles out a
generating function for the perfect fluid theory as a functional of the
corresponding distribution function, which determines completely the
dynamics, thus in particular its causal properties. We obtain a very
simple sufficient condition on the associated distribution function, to
ensure causality of the whole perfect fluid theory. We call the
theories satisfying these condition {\it manifestly causal theories}.
They include as particular case the Boltzmann, Bose, and Fermi gasses.

{}From the statistical non-dissipative divergence theory we can easily
define a {\it dissipative} divergence theory, keeping its statistical
origin. We shall say in this case that the dissipative theory thus
obtained is a {\it natural extension} of the non-dissipative one. This
extension, has an important property: If the original non-dissipative
theory is manifestly causal, then its dissipative extension is also
manifestly causal.  In particular, since the theories of equilibrium
Boltzmann, Bose, and Fermi gasses can be cast in the form of manifestly
causal non-dissipative theories, we conclude their natural extensions
are also manifestly causal, thus considerably generalizing and
simplifying of the works of \cite{s}, \cite{lmr}, \cite{ss},
\cite{nr1}. In spite of its naturality, this extension has some
drawbacks, as explained in Section II, it is only well defined for some
values of the dissipation variables, which for instance do not cover
any whole neighborhood of the local equilibrium submanifold. Thus, some
compatible, but otherwise arbitrary, extension has to be made in order
to cover the remaining values the dissipation variables can take.

Later on, in section III, we shall relax the statistical condition and
allow theories which have a {\it pseudo-statistical} origin, that is
where the three tensors fields, $N^a$, $T^{ab}$, and $A^{abc}$ are
averages over a future mass shell, but for different (although related)
distribution functions.  Again we shall find for this case a subclass
of  manifestly causal theories which includes the former one, and again
show that manifestly causal non-dissipative theories have natural
dissipative extensions which are also manifestly causal.

In section IV we shall discuss the expression for the entropies of this
theories.  The requirement of an entropy law is so strong that even
singles out a unique form of the entropy as a functional of the
generating function. The entropies we deal here are {\it dynamical
entropies} in the sense that they appear as further relations
(equations) between the fields as a consequence of its dynamics. We
find that, in contrast with the usual definitions of equilibrium
entropies, for the statistical theories they are linear functionals of
the distribution functions.  Nevertheless, when evaluated at given
equilibrium distribution functions, e.i. Boltzmann, Bose or Fermi
distribution functions, they coincide with the usual formulae coming
from sum-of-states considerations.

Finally, for completeness, we include two appendices, one is a short
review of the relativistic dissipative theories of divergence form, and
the other of relativistic statistical mechanics.

\section{Statistical Theories}

The divergence type fluid theories are briefly reviewed in Appendix A.
The simplest particular case of these theories is when the generating
function $\chi$ does not depend on dissipative variables, and therefore
is non-dissipative. In this case the theory describes a  perfect fluid,
with the particle-number current and the stress-energy tensor having
the usual structure, namely,
\begin{eqnarray*}
N^a &=& - \chi_{,\zeta \mu} \; u^a, \\
T^{ab} &=& \chi_{,\mu \mu} \; u^a u^b - \frac{\chi_{,\mu}}{\mu} \;
q^{ab}
\end{eqnarray*}
where $q^{ab} = g^{ab} + u^a u^b$, and $\zeta^a=\mu u^a$, with
$u^au_a=-1$.  So we have the particle-number of the fluid $n=-
\chi_{,\zeta \mu}$, the energy density $\rho = \chi_{,\mu \mu}$ and
pressure $p=- \frac{\chi_{,\mu}}{\mu}$ in terms of $\chi$.  It can be
seen that in this theory $\mu$ is one over the temperature of the
fluid, and $\zeta$ is a chemical potential per unit temperature.  The
choice $\chi = \chi(\zeta,\zeta^a)$ implies that the tensor $A^{abc}$
is identically zero, and this is consistent with the choice $I^{ab}=0$,
for the dissipation-source tensor.  In these theories the expression
for the entropy current given in the Appendix reduces to the usual one,
$S^a = n s u^a$, \hfil with $s=(\rho+p)/(nT) - \zeta$, and it satisfies
$\nabla_aS^a=0$.

We say that a perfect fluid theory is of {\it statistical type}, if its
generating function $\chi(\zeta,\zeta^a)$ can be written in the
following way
\begin{equation}
\chi(\zeta,\zeta^a) = \int f( \zeta + p_a \zeta^a) \; d\omega
\label{defchistat1}
\end{equation}
where the integral is on the future mass shell $p^ap_a=-m^2$, and $f$
is a smooth scalar function on the negative real line, which for large
(negative) values decays fast enought as to make the integral well
defined and $\chi(\zeta,\zeta^a)$ smooth in all its variables. Notice
that $\zeta < 0$, since it is a chemical potential, while $p_a\zeta^a <
0$, since it is the scalar product of two future directed time-like
vectors.

The simplest example is when $f(x) = k^2 e^{x/k}$,  where $k$ is
Boltzmann's constant, namely Boltzmann's gas.  Indeed, the expressions
for the particle-number current and the stress-energy tensors are in
this case given by
\begin{eqnarray*}
N_a &=&  \int p_a \; e^{(\zeta + p_a\zeta^a)/k} \; d\omega, \\
T_{ab} &=&  \int p_ap_b \; e^{(\zeta + p_a\zeta^a)/k} \; d\omega.
\end{eqnarray*}
That is, we interpret $N_a$ and $T_{ab}$ as the two first momenta of
the Boltzmann's equilibrium distribution function.

For an arbitrary statistical perfect fluid it is easy to see that the
$\chi$ defined as in (\ref{defchistat1}) corresponds to a statistical
theory with a distribution function given by $f^{\prime\prime}$.

An interesting property of this integral representation is that
causality is easy to analyze. Indeed following \cite{gl1} (see Appendix
A), we say that a perfect fluid theory is causal if:
$$
t_a E^a = \frac{1}{2} t_aM^a_{AB} \; Z^AZ^B < 0
$$
for all perturbations $Z^A = (\delta \zeta,\delta \zeta^a)$ and all
future-directed timelike vectors $t_a$. But for statistical perfect fluid
theories, using the expression of $\chi $ in terms of mass shell
integrals, this condition becomes:
\begin{equation}
t_a E^a = \frac{1}{2} \int (t_ap^a)
f^{\prime\prime\prime} \; \left(\delta \zeta + p_a
\delta \zeta^a \right)^2,
\label{energy1}
\end{equation}
for all perturbations $(\delta \zeta, \delta \zeta^a)$.  Thus we can
isolate from the set of all statistical theories a subclass of {\it
manifestly causal theories}, namely those having $f''' \neq 0$, $f'''
\geq 0$.~\footnote{Note that there are theories which are not
manifestly causal, but nevertheless are causal.} Notice that
Boltzmann's gas is inside this subclass, which also includes Bose and
Fermi's gases, since for them we have
$$
f^{\prime\prime}= \frac{1}{e^{-(\zeta+p_a\zeta^a)/k}
- \epsilon},
$$
(according of the value of $\epsilon$,  0, 1 or $-1$,  we have
respectively Boltzmann, Bose or Fermi gases), and it is easy to see
that for all of them $f^{\prime\prime\prime} \geq 0$. The generating
functions $\chi$ for Bose and Fermi distribution functions are obtained
by straightforward integration.

We now turn to the problem of extending these theories outside equilibrium.

The extension outside equilibrium we propose is the following:
Given a statistical perfect fluid theory characterized by a function
$f$, we define a {\it dissipative} divergence theory of statistical type by
the generating function
\begin{equation}
\chi(\zeta,\zeta^a,\zeta^{ab}) = \int f( \zeta + p_a \zeta^a + p_a p_b
\zeta^{ab}) \; d\omega.
\label{defchistat}
\end{equation}

Note that this extension is only valid for values of $\zeta^{ab}$ such
that the argument of $f$ is negative, while in general $\zeta^{ab}$ can
take values for which the argument is positive~\footnote{Near
equilibrium the theory should behave, at least in some respects, like
Eckart's theory. But in this theory there is a relation between
$\zeta^{ab}$ and the derivatives of the flow velocity which renders
totally unphysical any imposition on the sign of $\zeta^{ab}$.\hfill}.
Since $\zeta$ and $p_a\zeta^a$ are already negative for all $p_a$ in
the future mass shell, the values of $\zeta^{ab}$ for which the
argument is negative for all values of the momentum in the future mass
shell form a cone, $C^-=\{ \zeta^{ab} |\; \zeta^{ab}=\zeta^{ba}, \;
g_{ab}\zeta^{ab}=0, \; l_a l_b \zeta^{ab} \leq 0, \; \;\forall \;l_a \;
\mbox{null}\}$, of maximal dimension.  Thus this extension is not unique, the
equilibrium behavior of the statistical fluid only gives information
encoded in an $f$ which is only defined for negative values of the real
line, while here we seem to need an $f$ defined on the whole line.  But
the situation here is even worse, as shown in the following lemma.

\begin{lemma}
The function,
$$
F(c) \equiv \int_0^{\infty} f(-x+cx^2) x^2 dx,
$$
where $f$ is any smooth, positive definite function, which is of
compact support or decays faster than $x^{-4}$ is discontinuous at
$c=0$, having there a finite limiting value from the left and
diverging from the right.
\end{lemma}
{\bf Proof}

To prove that the limiting value from the left is finite just note that
if $c$ is negative, then the argument of $f$ just grows in absolute
value and so the decay assumptions on $f$ imply convergence. To prove
the statement about the the limiting value from the left, take a double
step function (a small square), $sc(x)$, smaller than $f$, then we
have,
$$
F(c) \geq G(c) \equiv \int_0^{\infty} sc(-x+cx^2) x^2 dx.
$$
So it suffices to prove the divergence on the limit for the function
$G(c)$. Assume for simplicity that $sc(x)$ is different from zero only in
the interval $[-1,1]$, and that there its value is $1$. For positive values
of $c$, the integral above has contributions from two intervals. If
$c$ is small, those intervals can be calculated up to first order in
$c$ obtaining $[1+c,0]$ and  $[\frac{1}{c}+1,\frac{1}{c}-1]$. The
contribution to the integral from the first one is finite for all
values of $c$, while the second can be estimated using the mean value
theorem to be bigger than $\frac{2}{c^2}+1+O(c)$ and so we see that it
diverges for $c$ going to zero from positive values.
\begin{flushright}
{\bf Q.E.D.}
\end{flushright}

This Lemma gives a strong argument against the possibility of extending
smoothly the definition of $\chi$ to values of $\zeta^{ab}$ such that
$p_ap_b\zeta^{ab} >0$, by simply extending the definition of $f$ to
positive values, no matter how smooth or how strong a decay condition
we impose on the extension.  But of course, there exist a lot of smooth
extensions of the generating function $\chi$ to the presently forbidden
values of $\zeta^{ab}$. Any of those, essentially arbitrary, smooth
extensions will be assumed to have been made in what follows, in
particular we require that extension in a neighborhood of equilibrium,
that is, in a neighborhood of the apex of the cone $C^-$. The results
on causality near equilibrium do not depend on the particular extension
chosen outside $C^-$.

For these statistically extended dissipative fluids theories also
a simple expression for the causality condition is easy to obtain, namely:
$$
t_a E^a = \frac{1}{2} \int (t_ap^a) f^{\prime\prime\prime} \left(\delta
\zeta + p_a \delta \zeta^a + p_a p_b \delta \zeta^{ab} \right)^2
$$
So we have have the following result:
\begin{theorem}
Let function $f: R^- \rightarrow R$ defining a statistical perfect
fluid be $C^3$, and such that the equilibrium theory is well defined.
If the equilibrium theory is manifestly causal, i.e.
$f^{\prime\prime\prime} > 0$, then the extended dissipative theory is
also causal in a neighborhood of equilibrium.
\end{theorem}

\noindent {\bf Proof}

The above expression shows causality for all values of $(\zeta,
\zeta^a, \zeta^{ab})$ such that $\zeta^{ab} \in C^-$. Since the cone is
of maximal dimension, partial derivatives along directions inside the
cone suffices to determine completely the differentials of $\chi$ at
the apex of the cone, that is at equilibrium. Thus the smooth extension
outside the cone can not change the causality properties of the
equilibrium configuration. The result extends to a neighborhood of
equilibrium trivially by noticing that in our setting causality is a
continuous property.
\begin{flushright}
{\bf Q.E.D.}
\end{flushright}
Particular examples of this theorem are the dissipative extensions
of Boltzmann, Bose and Fermi's gases, given by
$$
f^{\prime\prime}= \frac{1}{e^{-(\zeta+p_a\zeta^a+p_ap_b\zeta^{ab})/k}
- \epsilon}
$$
according of the value of $\epsilon$ is 0, 1 or $-1$ respectively.  In
all cases is easy to see that $f''' > 0$, so they are smooth and
manifestly causal for all values of the parameters for which the
integral expression converges, in and off equilibrium. This generalize
and simplify our previous work \cite{nr1}, which in turn was a generalization
of other results \cite{s}, \cite{ss}.

\section{Pseudostatistical Theories}

There is an even larger set of theories defined in terms of the
integral of certain function $f$ on a future mass shell, for which it
is straight forward to find sufficient conditions on $f$ such that the
resulting divergence theory is causal.  We say that a dissipative fluid
of divergence form is of {\it pseudo-statistical} origin, if its
generating function $\chi$ can be written as
$$
\chi(\zeta,\zeta^a,\zeta^{ab}) = \int
f(\zeta+p_ap_b\zeta^{ab},p_a\zeta^a) \; d\omega
$$
where the integral is on the future mass shell, and $f=f(x,y)$ is a
scalar function of two variables, $x \equiv \zeta +  p_ap_b\zeta^{ab}$
and $y \equiv p_a\zeta^a$.  We call them pseudo-statistical because of
the tensors $N^a$ and $A^{abc}$ can be thought as coming from a
distribution function $f_{xy}$, \footnote{Here the subindices indicate
partial derivatives with respect to the argument signaled by the
index.} while the stress-energy tensor $T^{ab}$ can be thought as
coming from a {\it different} distribution function, $f_{yy}$, in fact,
\begin{eqnarray*}
N_a &=& \int p_a \; f_{xy} \; d\omega \\
T_{ab} &=& \int p_ap_b \; f_{yy} \; d\omega \\
A_{abc} &=& \int p_a p_{<b}p^{c>} \; f_{xy} \; d\omega.
\end{eqnarray*}
Here the values of the variables $(\zeta,\zeta^a)$ have the same
restrictions as in section above.  The lemma 1 does not apply here and
there are cases, like Boltzmann, where one can obtain a
pseudo-statistical extension for all values of $\zeta^{ab}$ starting
from an equilibrium statistical theory.

The pseudo-statistical theories are interesting since again the
causality condition is simple and it is easy to impose a sufficient
condition such that the theory is manifestly causal. Indeed, it is easy
to see that in this case the causality condition is,
\begin{eqnarray*}
t_aE^a &=& \frac{1}{2}\; \int   \left(t_ap^a
\right)  \;\left[ f_{xxy} ( \delta\zeta )^2 +  f_{yyy} \left(p_b
\delta\zeta^b\right)^2 + f_{xxy}  \left(p_{<d}p_{e>} \delta\zeta^{de}\right)^2
+ \right.\\
&& \left. 2  f_{xyy} \delta\zeta \;  \left(p_b\delta\zeta^b\right) + 2
f_{xxy} \delta\zeta \; \left( p_{<d}p_{e>} \delta\zeta^{de}\right) + 2
f_{xyy} \left(p_b\delta\zeta^b\right) \; \left( p_{<d}p_{e>}
\delta\zeta^{de}\right) \right] \; d\omega.
\end{eqnarray*}
Rearranging terms we have
$$
t_aE^a=\frac{1}{2}\; \int  \left(t_ap^a \right) \;\left[
f_{xxy} \left( \delta\zeta + p_{<d}p_{e>}
\delta\zeta^{de}\right)^2 + f_{yyy} \left(p_b \delta\zeta^b\right)^2 +
2 f_{xyy} \left( \delta\zeta + p_{<d}p_{e>} \delta\zeta^{de}\right)\;
\left(p_b\delta\zeta^b\right) \right] \; d\omega,
$$
and defining $\delta x=\delta \zeta+ p_{<a}p_{b>}\delta\zeta^{ab}$ and
$\delta y=p_a\delta\zeta^a$,
\begin{equation}
t_aE^a=\frac{1}{2}\; \int   \left(t_ap^a \right)
\;\left[f_{xxy}(\delta x)^2 + f_{yyy} (\delta y)^2 + 2 f_{xyy} \;\delta
x \; \delta y \right] \; d\omega.
\label{energy2}
\end{equation}
Thus, if we assume $f_{xxy} \geq 0$, $f_{yyy} \geq 0$ and $(f_{xyy})^2
\leq f_{yyy} f_{xxy}$, we obtain,
$$
t_aE^a \leq \frac{1}{2}\; \int  \left(t_ap^a \right) \;  \left(
\sqrt{f_{xxy}}\; |\delta x| - \sqrt{f_{yyy}} \;| \delta y| \right)^2 \;
d\omega,
$$
and again isolate a subclass of dissipative
theories (containing the one previously described) which
are {\it manifestly causal}, namely those which satisfy,
$$
f_{xxy} \geq 0 \hspace{1cm} f_{yyy} \geq 0 \hspace{1cm} (f_{xyy})^2
\leq f_{yyy} f_{xxy}.
$$

We thus have an obvious generalization of the previous theorem.

\begin{theorem}
Let function $f: R^-\times R^- \rightarrow R$ defining a perfect
(equilibrium) pseudostatistical fluid be $C^3$, and such that the
equilibrium theory is well defined.  If the equilibrium theory is
manifestly causal, then the extended dissipative theory is also causal
in a neighborhood of equilibrium.
\end{theorem}

Remark:
There are causal $f$'s whose derivatives does not satisfies conditions
above. Even more, if we extend the theory in a pseudostatistical way
for positive arguments, then the decay properties would in general
imply the above condition can not be imposed, so even in this case,
causality for naturally extended theories seems to hold only in a
restricted domain of possible field variables.


The generating functions $\chi$ associated with a statistical or a
pseudo-statistical dissipative divergence type theories, can be thought
as particular solution of equation (\ref{chieq}) of Appendix B. This
equation is obtained by imposing some conditions on the divergence type
theories, related with the symmetry and trace of the third rank tensor,
$A^{abc}$, of the theories.  Those conditions come naturally from
kinetic theory and are explained in that Appendix. From the above
subclass of theories it seems then that there are globally defined, and
nicely behaved solutions to those equations, but they are not
statistical, as one would have liked them to be, they are at most
pseudostatistical.

\section{Entropy}

As seen in the introduction and Appendix A, the requirement of an
entropy law is a very strong condition which substantially
constraints  the kinematics and dynamics of the fluid theories, since
imposes that it can only depend on a freely given generating function,
$\chi$, (essentially an equation of state), and a --basically freely
given-- dissipation source tensor $I^{ab}$.  Even more, it gives a
fixed formula for the entropy as a functional of the arbitrary
generating function defining particular theories.  This entropy is of
dynamical origin and in principle it is not related to the entropy
concept coming from information theory as applied to equilibrium
configurations. In fact, the {\it entropy law} property can be divided
in two conditions the more stringent one is the existence of an extra
vector field whose divergence is, as a consequence of the other
equations, a local function of the dynamical field (and not of their
derivatives), and a weaker one which requires this last function to be
non-negative and to vanish at equilibrium. The first condition is the
one which restricts the theory, the one which gives its rigidity, and
is the one which determines completely the entropy functional, while
the other condition, in this picture, just puts mild conditions on the
dissipation-source tensor.  We shall see in what follows how these
conditions work for the natural extensions of statistical theories.

In the theories of divergence type the entropy is given by, (see
Appendix A.)
\begin{eqnarray}
S_a &=& \frac{\partial \chi}{\partial \zeta^a} - \zeta N_a - \zeta^b
T_{ab} - \zeta^{bc} A_{abc}  \nonumber \\
&=& \chi_a - \zeta^A \; \frac{\chi_a}{\partial \zeta^A},
\label{defS}
\end{eqnarray}
where $\zeta^A=(\zeta,\zeta^a\zeta^{ab})$.
For the naturally extended statistical theories this translate into, (See
Appendix B.)
$$
S_a = \int p_a \left( f^{\prime} - \zeta^A P_A \; f^{\prime\prime} \right)
\; d\omega
$$
where $P_A \equiv (1,p_a,p_ap_b)$.

We observe that this entropy is a {\it linear} functional of the
distribution function, and so there is no convexity property a this stage.
Its form does not depends on any {\it statistical counting of states}
assumption, once the distribution function $f''$ is given, the entropy is
determined.

The relation with the usual concepts and formulae comes about when one
identifies, via Boltzmann equation,  the tensor $I^{ab}$ with a collision
integral.
That is we pretend that the dependence of this tensor with $\zeta^A$ is
only throught the distribution function.
In that case, the weaker condition we had in the dynamical picture, i.e. the
vanishing of the divergence of the entropy current to second order in the
dissipation variables implies, via the collision functional, that the global
equilibrium states have the Gibbs form, which are
thus linked to the statistics considerations used to deduce the form of
that functional.
To see how this works we assume the above setting and compute the entropy for
the global equilibrium states of the usual statistics, (they enter only in the
choice of collision functional used to define $I^{ab}$), that is the examples
considered above.
Thus assume that the associated distribution function to the fluid theory
is of the form
$$
f^{\prime\prime} = F = \frac{\eta}{e^X - \epsilon},
$$
where $\eta$ is a constant proportional to $1/h^3$ with $h$ the Plank's
constant and for simplicity we denote $X = - P_A\zeta^A /k= -(x+y)/k
=-(\zeta + p_a\zeta^a + p_a p_b\zeta^{ab})/k$. It is direct to see
that
$$
f^{\prime} = -\frac{\eta k}{\epsilon} \left[ \ln \left( e^X -\epsilon \right)
- X \right].
$$
If we define $\Delta = 1+\epsilon \; F/\eta$, then we can write $e^X =
(\eta\Delta)/F$; and
$$
f^{\prime} = \frac{\eta k}{\epsilon} \; \ln(\Delta).
$$
So the entropy current density functional $S^a$ can be written as
\begin{eqnarray*}
S_a &=& \int p_a \left( f^{\prime} + k X \; f^{\prime\prime} \right) \;
d\omega\\
&=& k\int p_a \left( \frac{\eta}{\epsilon} \; \ln(\Delta) + \ln \left(
\frac{\eta\Delta}{F} \right) \; F \right) \; d\omega \\
&=& - k\int p_a \left( F \; \ln \left(\frac{F}{\eta}\right) -
\frac{\eta}{\epsilon} \; \Delta \; \ln (\Delta) \right) \; d\omega.
\end{eqnarray*}
The last expression correspond to the usual entropy functional defined
in relativistic kinetic theory whose equilibrium states corresponds to
Bose or Fermi's equilibrium distribution function if $\epsilon$ is 1 or
$-1$, respectively. The case $\epsilon =0$ corresponds to the Boltzmann's
distribution function, and is easy to see that $S^a$ given by
(\ref{defS}) has the usual form
$$
S_a = - k\int p_a  F \; \ln \left(\frac{F}{\eta}\right)\; d\omega
$$
Thus the usual formulae for the entropy functionals is recuperated, but
they are only valid for global equilibrium distribution functions, that
is the properties usually assigned to entropies are only relations
among certain distribution functions, and do not seems to hold outside
that set, even at local equilibrium states.

\section{Conclusions}

We have presented a variety of dissipative divergence type theories
with a statistical origin, in the sense that the tensors of the theory
can be expressed as appropriate functions of the three first momenta of
a suitable distribution function. This represents a relation with
kinetic theory, which is manifest in the integral expression for the
resulting generating function $\chi$. From this integral expression we
could easily derive a simple condition on the associated distribution
function for the resulting theory to be causal, even for some states
even far momentarily equilibrium states. This condition had been easily
verified for the natural extensions to non-equilibrium states of
equilibrium distribution functions associated with Boltzmann, Bose and
Fermi's statistics. The results obtained for these particular examples
represent a great simplification of our previous work \cite{nr1}, which
was a generalization of former works \cite{s}, \cite{ss}.

The dynamical entropy defined in the divergence type theories has not
been related, in principle, to the entropy concept coming from a
statistical theory as applied to equilibrium distribution functions.
This dynamical entropy is just a vector field (constructed from the
fluid fields) whose divergence is a pointwise function in the fields
and its definition does not use anything about equilibrium
configurations.  This expression has a surprising form, for in the
theories here considered it is linear on the associated distribution
function.  A remarkable fact for this dynamical entropy is that for
momentarily equilibrium configurations corresponding to the
distribution functions associated with the usual Boltzmann, Bose and
Fermi statistics; it takes the familiar form encounters in statistical
mechanics when applied to equilibrium configurations.  Thus, from this
point of view, the expressions usually used are only valid for a very
restricted set of distribution functions and when so, only for some
regions of the non-equilibrium configuration variables manifold, namely
that where the natural extension holds.

There exists a serious limitation for this integral representation of
the statistical theories, as has been shown with the Lemma 1. This
Lemma says that we can not extend smoothly the definition of $\chi$, to
values of $\zeta^{ab}$ such that $\zeta^{ab}p_ap_b > 0$, simply by
extending the definition of $f$ to such values. In other words, the
statistical interpretation of $\chi$ can not be extended. We can, of
course, extend smoothly the definition of $\chi$ to the forbidden
values of $\zeta^{ab}$ in an almost arbitrary way. All those extensions
will be, by continuity, causal in a neighborhood of the region where
$p_ap_b\zeta^{ab}$ is non-negative, in particular in a neighborhood of
equilibrium.  Does the limitation in the integral representation, have
any physical interpretation?. The condition $l_al_b\zeta^{ab} \leq 0,
\; \forall l^a$ null seems to be unphysical if we look near
equilibrium, for there the theory should resemble Eckart, and in that
theory it is known that there does not exist any restricting  condition
on the dissipative variable $\zeta^{ab}$.  So the problem, if this is
really a problem, should be in the fluid approximation, which in this
case is the pretention that we can describe the evolution of a
distribution function via Boltzmann's equation via the evolution of a
finite set of variables, namely $ \zeta^A = (\zeta, \zeta^a,
\zeta^{ab})$. While this seems to work very well at equilibrium, it
might be that it fails completely away from equilibrium~\footnote{It is
not clear in this context what it means expressions like {\sl near
equilibrium} and its relation with, for instance, the {\sl the size of
$\zeta^{ab}$}.}, and for an appropiate description an infinite of field
variables are needed. In that case a natural extension seems to be
possible, provided the distribution function is suitably extended for
positive values of its argument.  This solution is of the type, {\sl
add more dynamical fields} in order to get some sort of cutoff. Other
type of possible extension is to avandon the direct relation between
distribution momenta and dynamical variables outside equillibrium, but
still retain the flavor of a statistical theory in the sense of the
representation as integral over mass shells. This is easily implemented
by changing the argument of the distribution function in a non-linear
way, for instance:  $x=\zeta + p_a\zeta^a + p_a p_b\zeta^{ab} +
\lambda (p_a p_b\zeta^{ab})^2)$.  This solution is of the type, {\sl
add a constant}, in order to get a non-linear cutoff. Both sound
interesting and perhaps the solution lies in between.

It has been possible to extend the proof of causality given for the
statistical theories to a bigger set of divergence theories. For this
set, the three tensors fields, $N^a$, $T^{ab}$, and $A^{abc}$ are also
averages over a future mass shell, but for different (although related)
distribution functions.  We do not have any application for this larger
class, but believe it should be of relevance for describing some
physical phenomena.  In this case extensions seems to be possible for
all values of the non-equilibrium variables, but such extensions in
general are not causal for all such values, and so do not seem to be
very physical.  Are there special extensions which from a manifestly
causal equilibrium statistical theory yield a causal pseudo-statistical
non-equilibrium theory?


\section{Appendix A: Review of dissipative flu\-id theo\-ries of
di\-ver\-gen\-ce type}

Following  \cite{gl1},  \cite{p}, \cite{lmr}, we define a Dissipa\-tive
Fluid Theo\-ry of Di\-ver\-gence Type as a theory having the following
three properties: 1) The dynamical variables can be taken to be the
particle-number current $N^a$, and the (symmetric) stress-energy tensor
$T^{ab}$. 2) The dynamical equations are
\begin{eqnarray}
\nabla_a N^a &=& 0, \label{def11} \\
\nabla_a T^{ab} &=& 0, \label{def21} \\
\nabla_a A^{abc} &=& I^{bc}, \label{def31}
\end{eqnarray}
where the tensors $A^{abc}$ (tensor of fluxes) and $I^{bc}$
(dissipation-source tensor) are local functions of the dynamical
variables $N^a$ and  $T^{ab}$, and are trace free and symmetric in the
last two indices. 3) There exists and entropy current $S^a$ (local
function of $N^a$ and $T^{ab}$) which, as a consequence of the
dynamical equations, must satisfy
\begin{equation}
\nabla_a S^a = \sigma
\label{entrop}
\end{equation}
where $\sigma$ is some positive function of $N^a$ and $T^{ab}$.

It is proved in \cite{gl1}, \cite{p}, that a theory having these
three properties is determined by specifying a single scalar generating
function $\chi$ and the tensor $I^{ab}$ as functions of a new set of
dynamical variables $\zeta, \zeta^a, \zeta^{ab}$ (with the later trace
free and symmetric). The dynamical equations for these variables are
(\ref{def1}-\ref{def3}), with
\begin{eqnarray}
N_a &=& \frac{\partial^2 \chi}{\partial \zeta \partial \zeta^a}
\label{defn1} \\
T_{ab} &=& \frac{\partial^2 \chi}{\partial \zeta^a \partial \zeta^b}
\label{deft1} \\
A_{abc} &=& \frac{\partial^2 \chi}{\partial \zeta^a \partial \zeta^{bc}}
\label{defa1}
\end{eqnarray}
while the entropy current is given by
\begin{equation}
S_a = \frac{\partial \chi}{\partial \zeta^a} - \zeta N_a - \zeta^b
T_{ab} - \zeta^{bc} A_{abc}
\label{defs}
\end{equation}
while its divergence is
\begin{equation}
\sigma = - \zeta^{ab} I_{ab}.
\label{defsig}
\end{equation}

It is helpful to represent the collection of dynamical variables as,
$\zeta^A=(\zeta, \zeta^a, \zeta^{ab})$, and  the dissipation-source
tensor as, $I_A=(0,0,I_{ab})$. Equations (\ref{def1}-\ref{def3}) can
then be written in this notation as
\begin{equation}
M^a_{AB} \nabla_a \zeta^B = I_A \label{eq1}
\end{equation}
with
$$
M^a_{AB} = \frac{\partial^3 \chi}{\partial \zeta_a \partial \zeta^A
\partial \zeta^B}
$$
The system of equations (\ref{eq1}) is automatically symmetric,
since partial derivatives commute. We say that a symmetric system is
hyperbolic if the vector
$$
E^a \equiv \frac{1}{2} \;M^a_{AB} \; Z^AZ^B
$$
lies to the future of some space-like three-dimensional sub-space of
the tangent space, for all non-vanishing $Z^A$. The system is called
space-time causal if furthermore $E^a$ lies within the future light
cone (i.e., if $E^a$ is a future-directed timelike vector), for all
non-vanishing $Z^A$.  The property of hyperbolicity ensures that system
(\ref{eq1}) has a well-posed initial-value formulation, while
space-time causality ensures that no fluid signals can propagate faster
than light.  Because the underlining statistical mechanics origin of
fluids, it is natural to demand causality even for its dissipative
versions, this implies conditions on the generating function $\chi$.

Following \cite{gl1} we call a state, $\zeta^A$, i.e. the value of
$\zeta^A$ at a given time, a {\it momentarily equilibrium state}, if at
that time the entropy production vanishes, i.e. $\sigma =0$.  In the
generic case this implies $\zeta^{ab}=0$. i.e.  $\zeta^A_{me} =
(\zeta,\zeta^a,0)$. An {\it equilibrium state} or {\it strict
equilibrium state} is a time reversible state. In that case it can be
seen that generically not only $\zeta^A_e=(\zeta,\zeta^a,0)$, and
$I^{ab}(\zeta^A_e)=0$, but $\zeta$ is constant, and $\zeta^a$ is
Killing.  For these strictly equilibrium states the modulus of
$\zeta^a$, that is $\mu \equiv \sqrt{-\zeta^a\zeta_a}$, can be
associated with  one over the temperature of the fluid, and its
direction $u^a \equiv \zeta^a/\mu$ with the 4-velocity of the fluid.
The variable $\zeta$ can be associated with a chemical potential per
unit temperature of the fluid.

\section{Appendix B: The relation with Kinetic Theory}

We consider a distribution of identical particles in a given, fix,
spacetime.  The particles interact via short-range forces, idealized as
point collisions. A distribution function $F(x,p_a)$ is defined
\cite{is} by the statement that
$$
F \frac{p_a}{m} \; d \Sigma_a \; d \omega
$$
is the number of world-lines cutting an element of 3-surface
$d\Sigma_a$ and having 4-momenta $p_a$ which terminate on a cell of
3-area $d\omega$ on the mass shell $p_a p^a=-m^2$.
This distribution function is the solution of the relativistic
transport equation
$$
p^a \nabla_a F(x,p^d) =
{\cal C}(x,p^d)
$$
where the derivative is along a curve in phase space which is geodesic,
starting at the point where we evaluate it and having as tangent vector
$p^a$. ${\cal C}$ is a collision term defined requiring that
$$
{\cal C}(x,p_d) \frac{d \omega}{m} \sqrt{-g} d^4x
$$
be the number of particles in the momentum range $d\omega$ around $p_d$
which are created by collisions in the 4-volume $\sqrt{-g} d^4x$,
around the point $x$. Given an expression for the collision term, in
general a functional of $F$, Boltzmann's equation determines the
dynamics of $F$. This collision term can not be an arbitrary function
of $F$, it has to satisfy certain physical properties. Usually  the
following general properties are required:
\begin{enumerate}

\item The form of ${\cal C}$ is consistent with 4-momentum
and particle number conservation at collisions.

\item The collision term ${\cal C}$ yields a non-negative expression for
the entropy production.
\end{enumerate}
To express these two conditions in a formal way, it is necessary to
relate the information on the distribution function with macroscopic
quantities, such as particle-number current or the stress-energy
tensor. To do this it is useful to introduce, given a dsitribution
function, $F$ the set of moments associated to this distribution, which
is the following hierarchy of totally symmetric tensors
$$
J_{a_1 \cdots a_n} \equiv \int p_{a_1} \cdots p_{a_n} \; F
\; d \omega, \;\;\;\; n=0,...,\infty
$$
where the integral is on the future mass shell, $p^ap_a = -m^2$.

These moments are not independent quantities because they satisfy the
following relations
\begin{equation}
J_{a_1 \cdots a_j \cdots a_i \cdots a_n}g^{a_j a_i} = - m^2 J_{a_1
\cdots a_{n-2}} \hspace{1cm} i,j = 1 \cdots n \hspace{1cm} i \neq j.
\label{contr}
\end{equation}
By virtue of transport equation, the $(n+1)$ momentum satisfies,
\begin{equation}
\nabla_a J^a_{a_1 \cdots a_n} = {\bf I}_{a_1 \cdots a_n},
\label{mom1}
\end{equation}
with the source tensor ${\bf I}_{a_1 \cdots a_n}$ defined as,
$$
{\bf I}_{a_1 \cdots a_n} \equiv  \int p_{a_1} \cdots p_{a_n} \; {\cal
C} \; d \omega.
$$

If we identify the first momentum with the particle-number current, that
is $N_a=J_a$; and the second momentum with the stress-energy tensor of
the fluid, that is $T_{ab}=J_{ab}$; then condition 1) on the collision
term can be written as
\begin{equation}
\int {\cal C} \; d\omega =0 \hspace{1.5cm} \int p_a \; {\cal C} \;
d\omega =0
\label{condCi}
\end{equation}

The {\it statistical entropy current density} is defined as a functional
of $F$ as follows,\footnote{This statistical entropy is not the dynamical
entropy
we are using. They only coincide for equilibrium states.}
$$
\widehat{S}_a(x) \equiv - \frac{1}{m} \int \phi(F) \; p_a d\omega
\hspace{1cm} \mbox{with} \hspace{1cm} \phi(F)=\left( F \ln(h^3F) -
\frac{\omega}{\epsilon h^3} \Delta \ln(\Delta)\right)
$$
and
$$
\Delta (x,p_{a}) = 1 + \epsilon \frac{h^3}{\omega} F(x,p_a)
$$
where $h$ the Planck's constant, $\omega$ is the spin-weight  (number
of available states per quantum phase-cell) and $\epsilon$ is $1$
for Bosons and $-1$ for Fermions. Then condition 2) on the collision
term can be written as,
\begin{equation}
\nabla_a \widehat{S}^a=-\frac{1}{m} \int \phi^{\prime}(F) \; {\cal C}(F) \; d
\omega \; \geq 0 \hspace{1cm} \mbox{where} \hspace{1cm}
\phi^{'}(F)=\ln \left(\frac{h^3F}{\Delta}\right).
\label{condCii}
\end{equation}

Finally, {\it local equilibrium distributions (or states)}, which are
defined requiring that for them the entropy production vanishes, namely
$\nabla_aS^a=0$. By first property imposed on the collision term, a
sufficient condition is that for this $F$'s
$$
\phi^{\prime}=\frac{\zeta}{k}+\frac{1}{k} \zeta^a p_a.
$$
This conclude a brief review of the basics Kinetic Theory we need.


To relate Kinetic theory with the divergence type fluid theories it is
necesary to make some identifications between the basic thensor fields
of both theories.
 As we did above, is direct to associate the two first moments $J^a$
and $J^{ab}$ with the particle-number current $N^a$ and the
stress-energy tensor $T^{ab}$.  These relations do not imply any
restriction on the generating function $\chi$, at least from the
symmetry and number of linearly independent elements of these tensors.
 Such a restriction on $\chi$ comes if one whishes to relate the third
 momentum of a distribution, $J^{abc}$, with the tensor $A^{abc}$ of
the divergence type theories.  This is sugested because, in analogy
with the divergence type dissipative fluids the first three momenta
satisfy
\begin{eqnarray}
\nabla_a J^a &=& 0  \label{mom3} \\
\nabla_a J^{ab} &=& 0 \label{mom4}\\
\nabla_a J^{a\langle bc \rangle} &=& {\bf I}^{\langle bc \rangle} \label{mom5}
\end{eqnarray}
where the symbol $\langle \; \rangle$ means symmetrization and trace
free.  But this analogy is only superficial and involves a very
stringent assumption.  Indeed, the above momenta equations are not a
closed system of equations, for ${\bf I}^{ab}$ is not a function of some
finite number of variables --for instance the previous two momenta--,
but it is a function of the distribution function, which in general is
a functional of all momenta. Thus systems are analogous only for those
situations which can be described with a distribution function which is
to a certain degree of aproximation a function of only the first two
momenta. At present there is not convincing argument that this is ever
the case, nevertheless we proceed here with the analogy.  Since
$
J_{ab}{}^b = - m^2 J_a = - m^2 N_a
$,
while,
$
A_{ab}{}^b=0
$, and assuming that $I_{ab}={\bf I}_{ab}$, the only  possible relation
between these tensors such that both dynamical system are equivalent
is:
\begin{equation}
J_{abc}=A_{abc}-\frac{m^2}{4} N_a \; g_{bc}
\label{def3mom}
\end{equation}
Since $J_{abc}$ is totally symmetric, this condition  imposes
conditions in $A_{abc}$ and $N_a$ (such that the right hand side of
(\ref{def3mom}) be totally symmetric), which in turn impose conditions
on the generating function $\chi$. These conditions are equivalent to
the following system of equations for $\chi$
\begin{equation}
\frac{\partial^2 \chi}{\partial \zeta^{[a} \; \partial \zeta^{b]c}} -
\frac{m^2}{4} \frac{\partial^2 \chi}{\partial \zeta \; \partial
\zeta^{[a}}  g_{b]c}  = 0
\label{chieq}
\end{equation}
Thus, for this identification to make sense the generating function
must be a solution of (\ref{chieq}). If we obtain such a solution, we
can then describe a fluid by the corresponding divergence type theory,
which thus has a statistical origin as given by the above
identification, even away from equilibrium.  A family of solutions of
this system of equations is given by
\begin{equation}
\chi(\zeta,\zeta^a,\zeta^{ab})=\int  f(x,y) \; d\omega,
\label{chisolution}
\end{equation}
where $x=\zeta+p_ap_b\zeta^{ab}$, $y=p_a\zeta^a$, the integral is on
the all future-directed momenta $p_a$, and $f$ is assumed to behave
asymtotically in such a way that the integral converges.  With
subindices in $f$ indicating differentiation, we have
\begin{eqnarray*}
\frac{\partial^2  \chi}{\partial \zeta^{a}  \partial \zeta^{bc}} &=&
\int \; p_a \; p_{<b}p_{c>} \; f_{xy}  \; d\omega \\
&=& \int \; p_a  \left( p_{b}p_{c}  + \frac{m^2}{4} g_{bc}
\right)\; f_{xy} \; d\omega \\
&=&  \int \; p_a p_{b}p_{c} \; f_{xy} \; d\omega+ \frac{m^2}{4} g_{bc}
\frac{\partial^2 \chi}{\partial \zeta \partial \zeta^a}
\end{eqnarray*}
where $p_{<b}p_{c>}=p_{b}p_{c}  + \frac{m^2}{4} g_{bc}$ means
symmetrization and trace free. So have
$$
\frac{\partial^2  \chi}{\partial \zeta^{[a}  \partial \zeta^{b]c}} -
\frac{m^2}{4}\frac{\partial^2  \chi}{\partial \zeta  \partial
\zeta^{[a}}\; g_{b]c} =\int d\omega \;\; p_{[a} p_{b]}p_{c} \;
f_{xy} =0
$$
As we have seen in the section II, there seems to be that there are not
{\it pure statistical} solutions, i.e. solutions of a the type:
$f(x,y)= f(x+y)$, with the appropiate differentiablility conditions on
$\chi$ they generate. Thus the analogy discussed above seems to be
unjustified.


\end{document}